\documentclass{aastex}
\usepackage{spr-astr-addons}
\usepackage{url}

\RequirePackage{color}

\begin{document}
\title{Asteroseismic modelling of the solar-like star $\beta$ Hydri}
\shorttitle{Asteroseismic modelling of $\beta$ Hydri}
\shortauthors{Do\u{g}an et al.}
\author{G. Do\u{g}an}
\affil{Department of Physics and Astronomy, Aarhus University,
Denmark}
\author{I. M. Brand\~{a}o}
\affil{Centro de Astrof\'{\i}sica da Universidade do Porto,
Portugal} \affil{Departamento de Matem\'{a}tica Aplicada - Faculdade
de Ci\^{e}ncias da Universidade do Porto, Portugal}
\author{T. R. Bedding } \affil{School of Physics A29,
University of Sydney, Australia}
\author{J. Christensen-Dalsgaard }
\affil{Department of Physics and Astronomy, Aarhus University,
Denmark}
\author{M. S. Cunha }
\affil{Centro de Astrof\'{\i}sica da Universidade do Porto,
Portugal}
\author{H. Kjeldsen }
\affil{Department of Physics and Astronomy, Aarhus University,
Denmark} \and
\begin{abstract}
We present the results of modelling the subgiant star $\beta$ Hydri
using the seismic observational constraints. We have computed
several grids of stellar evolutionary tracks using Aarhus STellar
Evolution Code (ASTEC, Christensen-Dalsgaard, 2008a), with and
without helium diffusion and settling. For those models on each
track that are located at the observationally determined position of
$\beta$ Hydri in the HR diagram, we have calculated the oscillation
frequencies using Aarhus adiabatic pulsation package (ADIPLS,
Christensen-Dalsgaard, 2008b). Applying the near-surface corrections
to the calculated frequencies using the empirical law presented by
Kjeldsen et al. (2008), we have compared the corrected model
frequencies with the observed frequencies of the star. We show that
after correcting the frequencies for the near-surface effects, we
have a fairly good fit for both $l$=0 and $l$=2 frequencies. We also
have good agreement between the observed and calculated $l$=1 mode
frequencies although there is room for improvement in order to fit
all the observed mixed modes simultaneously.
\end{abstract}

\section{Introduction}
$\beta$ Hydri is a G2-type subgiant star exhibiting solar like
oscillations. It is often referred to as the future of the Sun due
to having parameters (Table 1) close to those of the Sun, in
addition to being at a later evolutionary stage than the Sun. Being
rather evolved, $\beta$ Hydri exhibits mixed modes in its observed
spectrum. This makes the star particularly interesting for
asteroseismic studies as mixed modes carry more information about
the core than do the regular p-mode frequencies.

\begin{table}[h]
\begin{center}
 \caption{Parameters of $\beta$ Hydri as given in
literature. The radius was derived from the interferometric angular
diameter (North et al. 2007) and the revised Hipparcos parallax (van
Leeuwen 2007).  The mass was then derived from the mean density from
asteroseismology.}
\begin{tabular}{@{}lll}
\hline
Parameter & Value & Reference \\
\hline $M/M_{\odot}$ & 1.085$\pm$0.028
 & Kjeldsen et al. 2008
\\
$R/R_{\odot}$ & 1.809$\pm$0.015 & Kjeldsen et al. 2008
\\
$L/L_{\odot}$ & 3.494$\pm$0.087
 &  Current work\\
$T_{\rm{eff}}$(K) & 5872$\pm$44 & North et al. 2007
\\
$\rm{[Fe/H]}$&-0.08$\pm$0.04
 & Santos et al. 2005
 \\
$\Delta\nu_{0}$($\mu$Hz)& 57.24$\pm$0.16 & Bedding et al. 2007\\
\hline
\end{tabular}
\end{center}
\end{table}

An extensive analysis of this star was done by Fernandes \& Monteiro
(2003), who emphasized the possibility of employing seismic
constraints to remove partially the parameter degeneracy that exists
when only the non-seismic observational constraints (such as
$T_{\rm{eff}}$ and luminosity) are used. Those constraints alone are
not enough, as one can obtain the same location of a stellar model
in the Hertzsprung-Russell (HR) diagram using different stellar
parameters (such as mass, $X$, $Y$, $Z$, etc.). Fernandes \&
Monteiro (2003) used the large frequency separation,
$\Delta\nu_{0}$, obtained by Bedding et al. (2001), to show how to
constrain the mass independently and also note the need for
individual frequencies to further constrain the age of the star
using the characteristics of the frequency spectrum that are related
to the stellar core, such as mixed modes.

The first preliminary comparison of the individual observed
frequencies and model frequencies was done by Di Mauro et al.
(2003). They presented a model with large and small frequency
separation that are within the limits derived from the observations
(Bedding et al. 2001). They showed that the match between the model
and observed frequencies are satisfactory except for the $l$=1
modes, some of which are affected by avoided crossings. This
emphasizes again the importance of accurate asteroseismic
observations and detailed analysis in order to evaluate the stellar
interiors.

Here, we use the latest asteroseismic observational constraints
(Bedding et al. 2007) consisting of individual frequencies including
some modes which are identified to be possible mixed modes. We
present the methods to search for a best model and the resulting
best fits within the parameter space of our survey.

\section{Methods}
We started calculating the grids of evolutionary tracks with a wide
range of parameters and large steps of increment. Analyzing the
first grid we have selected the best models, around which we have
computed denser grids. The initial parameters of the grids are given
in Table 2.
\begin{table*}[htbf]
\begin{center}
\footnotesize \caption{Parameters used to compute the evolutionary
tracks}
\begin{tabular}{@{}llll}
\tableline
Parameter & Grid 1 & Grid 2& Grid 3\\
\tableline $M/M_{\odot}$ & 1.04--1.10 (with steps of
0.01)&1.076--1.084 (with steps of 0.002) & 1.076--1.084 (with steps
of 0.002)
\\
$Z/X$ & 0.018--0.022 (with steps of 0.001)
&0.018--0.022 (with steps of 0.001)& 0.018--0.022 (with steps of 0.001)\\
$Y$ & 0.23, 0.27, 0.28, 0.30
 &0.276--0.284 (with steps of 0.002)& 0.276--0.284 (with steps of 0.002)\\
Mixing length & &&\\
parameter ($\alpha$)& 1.4--1.8 (with steps of 0.2)
&1.75--1.85 (with steps of 0.025) &1.75--1.85 (with steps of 0.025) \\
Diffusion \& & && \\
{gravitational settling}& None & None & He \\
 \tableline
\end{tabular}
\end{center}
\end{table*}
Diffusion and gravitational settling of helium are added in the
third grid, in which not all the tracks have been successfully
completed. We have carried on our analysis with the tracks that did
not have convergence problems.

On each track in the grids, we have selected the models having
parameters within the observational uncertainty limits. We have
calculated the oscillation frequencies of those models and compared
with the observations. For the purpose of comparison we used the
frequencies resulting from dual-site radial velocity observations
with HARPS and UCLES spectrographs (Bedding et al. 2007).

Before comparing the calculated frequencies with the observed ones
we have applied near-surface corrections to the calculated
frequencies. The correction is needed due to the fact that existing
stellar models fail to represent properly the near-surface layers of
the solar-like stars, where the turbulent convection takes place.
This affects the high frequencies most; thus the correction should
be applied in a way that low frequencies are much less affected.

The situation is the same for the Sun, and the difference between
observed and calculated frequencies is shown to be well approximated
by the empirical power law given by Kjeldsen et al. (2008) as
\begin{equation}
\nu_{\rm{obs}}(n)-\nu_{\rm{best}}(n)=a\left[\frac{\nu_{\rm{obs}}(n)}{\nu_{0}}\right]^b,
\end{equation}where $\nu_{\rm{obs}}$ are the observed $l$=0 frequencies with radial order $n$, $\nu_{\rm{best}}$ are the corresponding calculated frequencies of the best model, and $\nu_{0}$ is a constant frequency chosen to be the frequency corresponding to the peak power in the spectrum, which is taken as 1000$\mu$Hz for $\beta$ Hydri.
Kjeldsen et al. (2008) used the solar data and models to calibrate
the exponent $b$, which is calculated as 4.90 for the Sun. Using
this solar $b$ value, and calculating $a$ for each model, we have
applied the near-surface corrections to the model frequencies. The
corrections for mixed modes are probably less than for pure p modes,
as discussed by Kjeldsen et al. (2008), and so we have set them to
zero. We have then selected the best models performing a chi-square
minimization test for the goodness of the fit as
\begin{equation}
\chi^{2}=\frac{1}{N}\sum_{n,l}\left(\frac{\nu_{l}^{\rm{model}}(n)-\nu_{l}^{\rm{obs}}(n)}{\sigma(\nu_{l}^{\rm{obs}}(n))}\right)^{2},
\end{equation}where $N$ is the total number of modes included, $\nu_{l}^{\rm{obs}}(n)$, and $\nu_{l}^{\rm{model}}(n)$ are the observed frequencies, and the corrected model frequencies, respectively, for each spherical degree $l$
and the radial order $n$, and $\sigma$ represents the uncertainty in
the observed frequencies.

\section{Results}
Properties of our best models are given in Table 3, with the
corresponding so-called \'{e}chelle diagrams in Figures 1 and 2 (see
Christensen-Dalsgaard (2004) and references therein for the
explanation of an \'{e}chelle diagram). The increase in the
systematic difference between the observed and model frequencies
with increasing frequency can be seen in the left panels of the
\'{e}chelle diagrams.

\begin{table}[h]
\begin{center}
 \caption{Parameters of the best models}
\begin{tabular}{@{}lll}
\hline
Parameter & From grid 2 & From grid 3 \\
\hline
$M/M_{\odot}$ & 1.082 & 1.082\\
$Z$& 0.01346 & 0.01266 \\
$Y$ & 0.278 & 0.284 \\
$\alpha$ & 1.825 & 1.775 \\
Age (Gyr) & 6.447 & 5.712\\
$R/R_{\odot}$ & 1.814 & 1.806\\
$L/L_{\odot}$ & 3.444 & 3.473 \\
$T_{\rm{eff}}$(K) & 5844 & 5869 \\
$a$ & -2.76& -3.76\\
$\chi^{2}$ & 7.71 & 11.93\\
\hline
\end{tabular}
\end{center}
\end{table}

\renewcommand\floatpagefraction{.9}
\renewcommand\topfraction{.9}
\renewcommand\bottomfraction{.9}
\renewcommand\textfraction{.1}
\setcounter{totalnumber}{50} \setcounter{topnumber}{50}
\setcounter{bottomnumber}{50}
\begin{figure}[h]
\includegraphics [scale=0.29,angle=0]{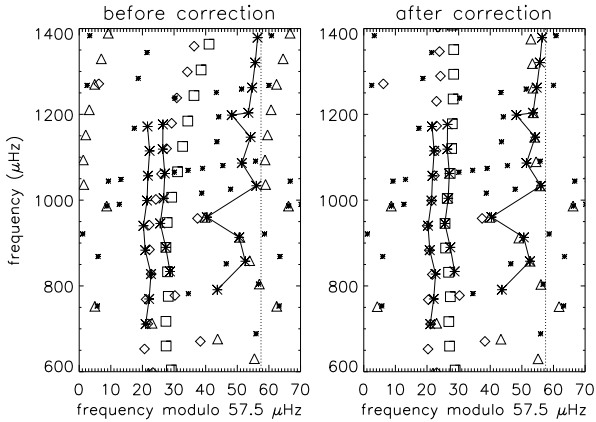}
\caption{\'{E}chelle diagram of the best model without gravitational
settling or diffusion. Left (right) panel shows the case before
(after) applying near-surface corrections. Stars denote the
observations, squares correspond to the model frequencies with
$l$=0, triangles to the model frequencies with $l$=1, and diamonds
to the model frequencies with $l$=2. Dotted vertical lines indicate
the value of the large frequency separation, $\Delta\nu_{0}$}
\end{figure}

The \'{e}chelle diagrams include all the observed frequencies, while
those having identified modes are connected by a solid line for
clarity. The observed frequencies which are not connected by a line
and shown with smaller symbols are the ones that are not assigned a
mode by the observers (Bedding et al. 2007); however, they note that
those frequencies will include some genuine modes although some of
them might be sidelobes or noise peaks. The size of the largest
symbols cover approximately 1.5-$\sigma$ uncertainty to both sides
on the horizontal axis. Three of the identified $l$=1 frequencies
that fall to the left side of the $l$=1 ridge are identified as
mixed modes by the observers (Bedding et al. 2007). The model
without He settling or diffusion reproduces the observed frequencies
very well except for the lowest mixed mode. It is seen from Fig.1
that some of the unidentified observed frequencies also match the
model frequencies quite well.
\begin{figure}[h]
\includegraphics [scale=0.29,angle=0]{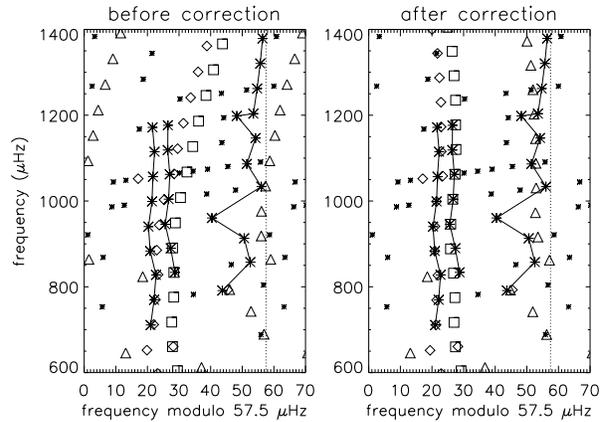}
\caption{\'{E}chelle diagram of the best model with He settling and
diffusion. The symbols are used in the same way as in Fig. 1} 
\end{figure}
When the diffusion and settling of helium is added, the agreement
between the model and the observations is less strong than the
previous case. Although the lowest mixed mode is reproduced in this
model, the mixed mode having the sharpest character (largest
departure from the $l$=1 ridge) is not. The highest mixed mode has
not been reproduced by any model frequency with $l$=1, suggesting
that $l$=3 frequencies might be investigated to search for a match
to that frequency. The positions of both models on the subgiant
branch are shown in Fig. 3.
\begin{figure}[h]
\includegraphics [scale=0.32,angle=90]{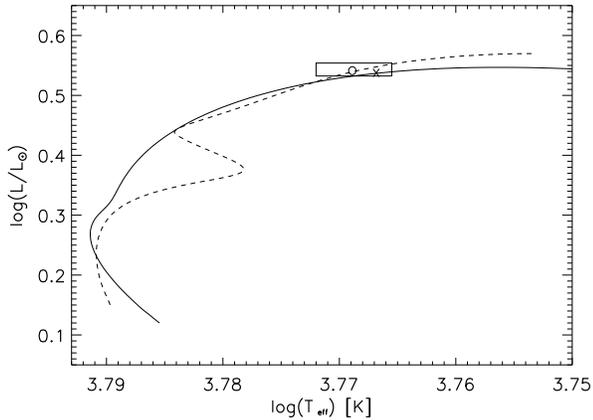}
\caption{Evolutionary tracks with He settling and diffusion (dashed curve, the best model marked with a circle) and without (solid curve, the best model marked with a cross) } 
\end{figure}
The model with diffusion is younger due to having completed the main
sequence phase faster, as the hydrogen mass fraction decreases
faster owing to diffusion and settling bringing helium to the core.
Furthermore, the hook shape in the evolutionary track with diffusion
is due to the fact that the star had grown a convective core up to
an extent where it contained almost 4\% of the stellar mass.

\section{Conclusions}
Our results justify that the empirical power law representing the
effect of near-surface layers in the Sun works for $\beta$ Hydri as
well. Our best models with and without He settling and diffusion
reproduce the observed $l$=0 and $l$=2 modes well; however, the fit
at $l$=1 modes is relatively poor due to the mixed modes, but still
satisfactory. It is important to investigate the possibility of any
of the observed modes being an $l$=3 mode.

Furthermore, our results are in agreement with the findings of
Fernandes \& Monteiro (2003), who derived, through the HR diagram
analysis, the mass to be 1.10$^{+0.04}_{-0.07}$ $M_{\odot}$, the
helium abundance to be between 0.25 and 0.30, and the stellar age to
be between 6.4 and 7.1 Gyr. We, however, note that the age of our
model with helium diffusion is less than the lower limit of the
cited result. Employing $\Delta\nu_{0}$, they found the mass to be
1.09$\pm$0.22 $M_{\odot}$, noting that the large uncertainty is to
be improved with the improved accuracy of the observations. Further
analysis may be carried out to investigate the effect of convective
core overshooting as our models had convective cores at some earlier
stage in their evolution.

\acknowledgments GD would like to thank Travis S. Metcalfe for
helpful discussions regarding the analysis. This research has been
partially supported by the Danish Natural Science Research Council.
IMB acknowledges the support by the grant SFRH / BD / 41213 / 2007
from FCT/MCTES, Portugal.

\makeatletter
\let\clear@thebibliography@page=\relax
\makeatother

\end{document}